\documentclass{imsart}

\RequirePackage{amsthm,amsmath,amsfonts,amssymb}
\RequirePackage[authoryear]{natbib}
\RequirePackage{url}
\RequirePackage[boxed]{algorithm2e}

\RequirePackage{graphicx}
\usepackage{soul}
\usepackage{epigraph}

\startlocaldefs

\endlocaldefs

\begin{document}

\begin{frontmatter}
\title{Markov Chain Monte Carlo for generating ranked textual data}

\runtitle{Markov Chain Monte Carlo for generating ranked textual data}

\begin{aug}

\author[A,B,C]{\fnms{Roy} \snm{Cerqueti}\ead[label=e1]{roy.cerqueti@uniroma1.it}}
\author[B]{\fnms{Valerio} \snm{Ficcadenti}\ead[label=e2]{ficcadv2@lsbu.ac.uk}},
\author[D]{\fnms{Gurjeet} \snm{Dhesi}\ead[label=e3]{dhesig74@gmail.com}}
\and
\author[D,E,F]{\fnms{Marcel} \snm{Ausloos}\ead[label=e4]{ma683@le.ac.uk}}

\address[A]{Sapienza University of Rome, Department of Social and Economic Sciences -- Piazzale Aldo Moro, 5 - I-00185, Rome, Italy. \printead{e1}}

\address[B]{London South Bank University, School of Business -- 103 Borough Rd, SE1 0AA, London, UK. \printead{e2}}

\address[C]{University of Angers, GRANEM -- SFR CONFLUENCES, F-49000 Angers, France.}

\address[D]{Group of Researchers for Applications of Physics in Economy and Sociology (GRAPES) -- 483 Rue de la Belle Jardiniere, B-4031 Liege, Belgium, \printead{e3}}

\address[E]{Bucharest University of Economic Studies, Department of Statistics and Econometrics -- 010374 Bucharest, Romania.}

\address[F]{University of Leicester, School of Business -- Brookfield Campus, Leicester LE2 1RQ, UK. \printead{e4}}
\end{aug}

\begin{abstract}
This paper faces a central theme in applied statistics and information science, which is the assessment of the stochastic structure of rank-size laws in text analysis. We consider the words in a corpus by ranking them on the basis of their frequencies in descending order. The starting point is that the ranked data generated in linguistic contexts can be viewed as the realisations of a discrete states Markov chain, whose stationary distribution behaves according to a discretisation of the best fitted rank-size law. The employed methodological toolkit is Markov Chain Monte Carlo, specifically referring to the Metropolis-Hastings algorithm. The theoretical framework is applied to the rank-size analysis of the hapax legomena occurring in the speeches of the US Presidents. We offer a large number of statistical tests leading to the consistency of our methodological proposal. To pursue our scopes, we also offer arguments supporting that hapaxes are rare (``extreme") events resulting from memory-less-like processes. Moreover, we show that the considered sample has the stochastic structure of a Markov chain of order one. Importantly, we discuss the versatility of the method, which is considered suitable for deducing similar outcomes for other applied science contexts.
\end{abstract}


\end{frontmatter}

\section{Introduction}
The rank-size analysis has a well-established theoretical framework in the broad context of applied science, whose grounding researches are often identified in the landmark contributions of Zipf and Mandelbrot, \cite{zipf1935psycho,zipf1949human,mandelbrot}.

Such a theory' quantitative ranked data behaviour through best-fit procedures. If the best fit procedures lead to statistically sounding results -- i.e., the goodness-of-fit indicators associated with the best fit curve have satisfactory values -- then one can reasonably infer that the best fit curve represents a macroscopic global unified system. Such a framework effectively summarises the considered microscopic local disaggregated data.

Getting hints from disaggregated data about a unified system can be crucial. Indeed, interpreting the best fit calibrated parameters might give several insights into the phenomenon under investigation. In this respect, the role of the family of parametric curves to be fitted is of particular relevance. 

For all these reasons, the rank-size theory is still fresh and constantly at the centre of the interest of scientists from different fields, ranging from economics \citep[see, e.g.][]{ACphysA, dimitrova, giesen} to the measurement of science like bibliometric studies \citep[see, e.g.][]{ausloos2013, auslooshapax} and seismology \citep[see, e.g.][]{fc}. Recent examples are also \cite{cerqueti2022combining} -- where the new death per million due to COVID-19 are ranked for several countries and fitted in a rank-size relationship to feed a clustering algorithm. Similarly, in \cite{ficcadenti2022rank}, the rank-size relationship is used to compare the team's final score in the Italian ``Serie  A'' football league. 

Usually, the rank-size theory is grounded on the exploration of empirical samples. However, the probabilistic structure of the rank-size analysis represents a challenging task to be discussed. Indeed, the probabilistic structure of a rank-size analysis is the stochastic process taking values in the set of the ranks and such that the rank-size distribution is the long-run distribution of the process outcomes. Such a structure provides relevant information. It allows us to assess how the ranked terms (in our context, the ranked hapaxes) occur; for the specific case of hapaxes, we observe that the stochastic process related to their occurrence is a Markov chain of order one. The knowledge of the stochastic structure leads then to the possibility of explaining the patterns of the rank-size distribution creation and the possibility of deriving future outcomes. 

Our paper aims to contribute to this direction in the paradigmatic context of linguistics and textual data. We aim to detect the stochastic structure of text-based rank-size laws. Specifically, we build a stochastic process whose evolution leads to reproducing the considered ranked data distribution. We seek a stochastic process whose asymptotic distribution reproduces a discretisation of the best fit curve.

Even if the scientific problem is rather general, the nature of the involved stochastic processes must be tailored to the linguistic nature of the given rank-size problem. In fact, the \textit{a priori} selection of a particular family of stochastic processes should follow and be in agreement with the features of the phenomenon under scrutiny. Despite its relevance, the general connection between the stochastic structure of a group of ranked data and the phenomenon measured through them seems, to us, not clearly debated in the literature \citep[studies like][certify the interesting ongoing debate about it]{Dodds2017}.
The approach to constructing a probabilistic model related to a rank-size law is grounded on interpreting the resulting rank-size distribution as an outcome of a stochastic process. In this respect, it is worth mentioning the preferential attachment context, where the idea is to define a step-wise procedure in the framework of the urn problem, as in Polya's process \citep[see][]{mamud} and in the presence of rules stating the addition of balls in the urn at every step. An example is in \cite{ausloos}, where a rank-size law was also discovered.
Also, in the presented context, a preferential attachment approach might be meaningful and valid -- under the obvious requirement that the asymptotic distribution of the stochastic process represents a statistically significant approximation of the rank-size law. Thus, to explore the proposed problem, we here present the specific rank-size setting coming from the text analysis environment by analysing a collection of rare words whose frequencies in a corpus correspond to the sizes in a rank-size analysis \citep[see][for a discussion on the text-length dependence of word-frequency distributions]{Corral2017}. The considered framework offers room for discussions and scientific explorations \citep[see, e.g.][]{ausloos2010punctuation, rovenchak}.

In this special context, as we will see below, we work with Markov chains with discrete states, representing a good choice for data-generating processes.

In particular, we move from the step done in \cite{JOI, cerquetieswa} and consider the rank-size analysis of the hapax legomena or, simply, hapaxes -- which are the words occurring only once in a given text; they can be viewed as rare events, in a text time series -- of an extensive collection of the speeches of the US Presidents. In \cite{cerquetieswa}, words from each speech are ranked in decreasing order, and their frequencies give the sizes to the rank-size relationship there used. In \cite{JOI}, only the hapaxes legomena are considered in each speech, so the hapaxes' size is the number of speeches in which they have been pronounced only once. The authors selected the Zipf-Mandelbrot law as the rank-size curve, and their best fit procedure leads to statistically satisfactory results \citep[see][for further details on the rank-size analysis]{JOI}. Importantly, in using the hapaxes of the corpus presented in \cite{JOI, cerquetieswa}, we provide and model a feature of a set of speeches produced by different writers over a long period. The corpus related to the US Presidents' speeches is of high quality for its inner characteristics of being composed of carefully written texts. Moreover, it is much more informative at a global level than the disaggregation of the individual speeches. These arguments let the considered example be of particular interest.

In such a rank-size context of text mining type, we propose the construction of a stochastic model that captures the process of leading to hapaxes inclusion in the speeches. With this aim, hapaxes are evaluated through their ranks and are assumed to be the result of sampling from a stochastic process producing rare events. We here discuss how Markov chains can be viewed as hapaxes generators \citep[see][for a broad description of the Markov chains]{norris}. In particular, we assume that the calibrated Zipf-Mandelbrot law found in \cite{JOI} drives the construction of the stationary distribution of such Markov chains. Detailed reasons to consider the Markov chain assumption as a suitable choice in our context are also provided. Moreover, we present statistical evidence that a Markov chain of order one fits the considered sample. The problem is discussed from a theoretical point of view, and later it is also assessed in the numerical framework of the Markov Chain Monte Carlo (MCMC) through a Metropolis-Hastings algorithm \citep[see, e.g. ][]{metropolis, hastings}.

The versatility of the MCMC approach is witnessed by several scientific papers in a large set of areas and scientific domains. In \cite{liuli}, the authors discuss merging Bayesian analysis and MCMC to overcome the computational complexity of the estimation procedures. In a more applied context, \cite{kwon} proposes an MCMC approach for building a robust visual tracker by empirically showing that the introduced device outperforms several visual tracking methods. In \cite{cerquetigiacalone}, there is a specific reference to MCMC for implementing forecasting of the volatility in the environment of financial markets. A list of relevant contributions in the literature highlighting the worthiness of MCMC for applications and for methodological advancements should include e.g., \cite{yang, zanella, austad2007parallel, luengo2020survey,mira2001metropolis,martino2018review}. 
The interested reader is also addressed to the high-level scientific contributions of Diaconis, whose reflections are synthesised in \cite{Diaconis9, Diaconis13}.

However, to the best of our knowledge, this is the first paper dealing with the derivation of the stochastic structure of the text-based rank-size laws through an MCMC approach.

The obtained findings are particularly interesting. They confirm that hapaxes occurrences in the individual texts collected in a corpus can be viewed as realisations of Markov chains on a sequential basis. Furthermore, the algorithmic construction of such a stochastic process is provided.
We notice that the Markov chain presented in this paper falls in the category of the stochastic processes, which have strong motivations for being used, and it generates pretty perfectly the rank-size law (see Section \ref{Markovass} for a discussion on this point). In this respect, we pay particular attention to the statistical aspect of the convergence of the MCMC algorithm -- see Section \ref{subsec:convergence}. Furthermore, we provide several tests for stating the worthiness of identifying the Markov chain as a generator of the rank-size law related to hapaxes. For these reasons, we point out that the proposed Markovian process identifies the stochastic structure of the considered rank-size law in a very general sense. The proposed method implies that several linguistic-based rank-size laws can benefit from our considerations of the role of underlying Markov processes. The case of the hapaxes of the US Presidents' speeches represents a pertinent illustration with a high degree of complexity. 

The rest of the paper is organised as follows. Section \ref{JOI} is devoted to the summary of the rank-size analysis presented in \cite{JOI}. It outlines the data, the methodology and the results of the rank-size analysis implemented in the quoted paper. Section \ref{Markovass} proposes a discussion of the motivations for considering that a Markov chain generates hapaxes by giving statistical evidence on the Markovianity of the ranked data. Section \ref{IIInew} contains a statistically rigorous analysis of the nature of the Markov chain of order one of the stochastic process generating the considered ranked data. Section \ref{stoch} is devoted to the formal construction of the Markov chain generating the ranked hapaxes; it also presents the convergence of the procedure -- hence further validating the proposed framework -- and a discussion of the obtained results. The last Section offers some conclusive remarks.

\section{Outline of the text-based rank-size context: the case of the hapaxes of the US Presidents' speeches}
\label{JOI}
This section presents a summary of the rank-size analysis proposed in \cite{JOI}.
Please refer to the quoted paper for all the technical details on finding hapaxes. 

The considered hapaxes list is taken from a corpus of 951 official speeches delivered by the US Presidents, ranging from 1789 (George Washington) to 2017 (Donald Trump). The transcripts of such speeches have been retrieved from the Miller Center website. The links to the talks\footnote{Notice that tweets and speeches before the pandemic are excluded} can be found in \url{https://millercenter.org/the-presidency/presidential-speeches}.

After some data treatment phases \citep[see][]{JOI, cerquetieswa}, 951 double columns are obtained  -- one for each speech -- they contain the words and their frequencies in the considered speech. The words with unitary frequency are the hapax legomena used in each speech. The resulting list of words said once (with repetitions) contains 509138 elements. \\
The distinct hapaxes have been collected in a unique set so that the hapaxes' occurrences in the corpus correspond to the number of speeches in which they have been employed as hapaxes. Such numbers are intuitively labelled as \textit{frequencies of the hapaxes}. The frequency of an hapax represents its size in our rank-size context. \\
Frequencies range from 1 to 250, so the number of distinct hapaxes is 31074. To be self-contained and to assist the reader in grasping the quantitative features of the hapaxes frequencies, we reproduce here the statistical summary of the hapaxes frequencies as reported in \cite{JOI} (see Table \ref{stat}).

\begin{table}
 \centering
 \begin{tabular}{|c|c|}
   \hline \hline
   \textbf{Statistical indicator} &    \textbf{Value} \\ \hline
N. OF DATA POINTS (HAPAXES) &  31074 \\
 MEAN ($\mu$)  & 16.3850 \\
 VARIANCE ($\sigma^2$) &   1034.2965 \\
STANDARD
DEVIATION ($\sigma$) & 32.1605 \\
 SKEWNESS  &  3.2451 \\
 KURTOSIS &
11.5989 \\
MEDIAN ($m$) & 3 \\
MAX & 250 \\
MIN & 1 \\
RMS  &   36.0934 \\
STANDARD ERROR & 0.1824 \\ \hline
$\mu/\sigma$ & 0.5095 \\
$3(\mu-m)/\sigma$ &  1.2486\\
\hline \hline
\end{tabular}
\caption{Main statistical indicators related to the frequencies of the hapaxes found in the considered speeches of the US Presidents.}\label{stat}
  \end{table}

Hapaxes are ranked in decreasing order; the hapax legomenon with frequency 250 has rank 1. We denote the size of an hapax by $s$ and its rank by $r$.

The best fit procedure is implemented by using a Zipf-Mandelbrot law, whose parametric functional shape is the following:
\begin{equation}
\label{zipf} s=f(r)=\frac{\alpha}{(\beta+r)^\gamma},
\end{equation}
where $\alpha, \beta, \gamma$ are real parameters; they have to be calibrated according to the considered dataset.

Table \ref{Table1} contains the results of the best-fit analysis, with the calibrated parameters and the goodness-of-fit indicators. As already observed in \cite{JOI}, there is excellent compliance of the ranked hapaxes with the curve presented in Eq. (\ref{zipf}) (see also Figure \ref{Fig1}).

 \begin{table}
 \centering
 \begin{tabular}{|c|c|c|}
   \hline \hline
$\hat{\alpha}$ & $\hat{\beta}$ & $\hat{\gamma}$\\ \hline $6.029
\times 10^8$ &        2540 & 1.896  \\ \hline
($5.676\times 10^8$, $6.381\times 10^8$) &  (2525, 2554) &(1.890, 1.902) \\
\hline \hline
\end{tabular}
\caption{Values of the calibrated parameters in the best-fit procedure, by using the Zipf-Mandelbrot law in Eq. (\ref{zipf}). In the brackets, the ranges of the confidence intervals at the level 95\%.}\label{Table1}
  \end{table}

\begin{figure}
\includegraphics[width=406.0pt]{MCMCfig.jpg}
\caption{Graphical representation of the best-fit curve according to formula (\ref{zipf}), with parameters given in Table \ref{Table1}. The scatter plot of the
considered dataset is also juxtaposed.}\label{Fig1}
\end{figure}

\section{Motivations behind the Markovian hypothesis on hapaxes generation}
\label{Markovass}

We provide here detailed arguments leading to interesting
connections between the proposed first-order Markov chains and the hapaxes pronounced by the US Presidents. In doing so, we justify and support the selection of Markov chains as stochastic processes to be used as a generator of the ranked data. In the next section, we validate the selection of the Markov chain of order one as the data generating process of ranked data, considering the case of the US Presidents' speeches hapaxes.
\begin{itemize}
\item Under the thematic point of view, there is a remarkable heterogeneity among the speeches of the US Presidents. Indeed, socio-historical contexts vary rapidly, and the US President has to timely intervene precisely, including in his discourses the most recent and relevant aspects of sociopolitical issues. Under the viewpoint of speeches' structures and the use of words, which we explore here, heterogeneity disappears, and the discourses exhibit a high degree of homogeneity. As the nature of the US Presidents' speeches suggests, the hapaxes meet some communication targets. Therefore, one can argue that the speeches cannot be viewed as disaggregated but rather compounded in a unified flow.
\newline
In analysing the sequences of speeches, a ``time unit" is the ``distance" between the appearance of two consecutive hapaxes, so that the first hapax occurs at time $t=0$ (initial value), the second hapax appears at time $t=1$, the third one is at time $t=2$ and so on. Two consecutive hapaxes may belong to the same speech; alternatively, one is the last of one speech, and the other is the first hapax of the following speech containing at least one hapax. We can conclude that the stochastic process that generates the hapaxes has a discrete-time underlying structure; this is perfectly in line with the considered Markov chain concept.

\item In general, Markov chains represent a useful device to model the stochastic structure of a data sample \citep[see, e.g.][]{shayeganfar09}. Importantly, by taking Markov chains as hapaxes generators, we align with a relevant and well-established strand of literature. In fact, the generation of words through Markov processes has been widely explored in many contributions \citep[see, e.g.][]{brainerd, newman}.
\newline
In this respect, we here mention with special attention \cite{nicolis}, where the authors discuss how deterministic chaos
may induce the construction of a Markovian stochastic process with finite states and whose stationary distribution provides a prefixed
distribution of ranked words. The difference between our approach and the quoted paper lies in the elements of the state space of the
Markov chains. \cite{nicolis} rank the words through their length and identify states as single characters; terms of length $L+1$ may then be viewed as ordered combinations of $L+1$ states of the chain so that the related Markov process is assumed to have order $L$. The separator between two words is considered to be a special character, the blank space. The quoted paper takes Zip's law as the prefixed distribution of the ranked words. 
\newline In our paper, we are close to \cite{nicolis}. Indeed, the states are the ranks of the hapaxes in our context. We are implicitly assuming that each hapax is one single symbol of a new ``alphabet", and hapaxes with the same rank -- frequency, for us -- are identified as a unique ``character". The separator between two hapaxes is the entire text between them.
Briefly, under the perspective advanced by \cite{nicolis}, we are constructing a novel alphabet, with several symbols given by the maximum considered rank $\bar{r}$ and words of length $L+1$ given by ordered sequences of hapaxes. We denote it by \textit{hapax language}. 
\newline
In the flow of the US Presidents' speeches, it results that one should search for a stochastic process generating words of length
$L+1$ in the framework of the hapax language. With this aim, we adopt an approach based on Markov chains of order $L$. We consider a
Zipf-Mandelbrot's law as target distribution of the ranked words.

\item The selection of the order $L$ of the Markov chain in this linguistic context is a relevant problem \citep[see, e.g.][]{brainerd,newman}. In \cite{begleiter}, one can find a thoughtful discussion on the varying order of a Markov chain for words generation in the light of the varying lengths of the objected words.
\newline
In our context, we check $L=1$ (see the next section). This outcome is expected in that it is grounded on an important reason.
\newline
Specifically, we recall that a first-order Markov process is a memory-less process. Thus, in the context of probability, the Markovianity of order one is associated with the memorylessness property of a stochastic process. Indeed, the probability of becoming one of the states of the chain in the next step depends only on the present state, and it is independent of the stochastic process' past states (see, e.g., \citealt[][]{Gudivada}). Thus, such a property represents the stochasticity of some phenomenon evolution. We argue that hapaxes in a President's speech are to be stochastically selected by the speaker (or his/her speechwriter). Indeed the hapaxes occur once in a speech; they are very specific to one speech and for some presidents. Of course, these words can be reproduced in other speeches by the same President or another. However, the words cannot appear twice in the same speech. Thus, one hapax can be considered to come from some \textit{dictionary black box} not containing others more often used words in the specific speech \citep[a similar idea can be found in the \textit{mental lexicon} described in][]{Allahverdyan2013}. An hapax in a given speech has no apparent relation to the other words in the speech (except for the grammatical obligations). In this respect, the dictionary black box has an ``infinite memory" of the speeches (even of the words to come). However, this does not contradict the picking up of a to-be-hapax word. Thus, an infinite number of steps has been taken into account in constructing the dictionary black box; but the chosen word is entirely stochastic and so-called memory-less.

It is worth noting that, beyond what is presented here for hapaxes, we are aware that several memory steps can occur between two identical words. Indeed, there is clear evidence of the positive appeal of introducing a large variety of words in writing \citep[see, e.g.,][]{kadhim2022lexical}. Thus, looking at correlations between hapaxes in different speeches might be interesting. Surely, such a research theme has practical meaning and allows advancing the methodology of stochastic processes and operational research. Indeed, such a memory inclusion might lead to a fractional calculus approach, as one of the authors of the present paper did, for example, \cite{ebadi}. However, this is outside the current logistics approach and quite outside the present investigation. Of interest, no doubt.

\end{itemize}

\section{Statistical evidence of the Markovianity of the stochastic process for hapaxes generation}
\label{IIInew}

We here statistically discuss the nature of the stochastic processes that serve as hapaxes generators according to the considered dataset.

All the random quantities will be contained in a probability space
$(\Omega, \mathcal{F}, \mathbb{P})$.

We start from an empirical sample made by 509138 consecutive observations; the elements of such a sequence are intuitively labelled by successive integers, representing the discrete time of the stochastic process. The observations are the ranks of the hapaxes, as they appear at the end of the observation phase. According to this perspective and in line with the ranks of hapaxes in the considered corpus, the 509138 observations are integer numbers taken from the set $\mathbf{Rank} = \{1,\dots, 227\}$. {For example, in the \textit{``First Inaugural Address''}, April 30, 1789, one finds the following:}

\epigraph{\textit{Fellow Citizens of the \textbf{Senate} and the House of Representatives:}

\textit{\textbf{Among} the \textbf{vicissitudes} \textbf{incident} to \textbf{life}, no event could have \textbf{filled} me with \textbf{greater} \textbf{anxieties} than that of which the \textbf{notification} was \textbf{transmitted} by your order, and \textbf{received} on the \textbf{fourteenth} day of the present \textbf{month.}}}{\textit{George Washington}}

Those hapaxes highlighted in bold are substituted by their corresponding rank in the ranked list of hapaxes so that it can be seen as:

\epigraph{\textit{Fellow Citizens of the \textbf{43} and the House of Representatives:}

\textit{\textbf{12} the \textbf{206} \textbf{150} to \textbf{54}, no event could have \textbf{143} me with \textbf{41} \textbf{215} than that of which the \textbf{207} was \textbf{154} by your order, and \textbf{64} on the \textbf{214} day of the present \textbf{98.}}}{\textit{George Washington}}

The problem of hapaxes generation can be synthesised as follows.

\begin{itemize}
\item[\textbf{Problem P}]
\textit{If a President has pronounced a hapax -- whose rank in the whole set of Presidents' speeches is $i \in \mathbf{Rank}$ -- which is the
probability that the consecutive hapax has rank $j \in
\mathbf{Rank}$?}
\end{itemize}

\textbf{Problem P} relies on the data generating process of the ranked data. According to the arguments reported in Section \ref{Markovass}, we here provide statistical evidence that such a process is a Markov chain of order one, with $\mathbf{Rank}$ as states space. To check the Markovianity of the ranked data, \textbf{Problem P} is here faced only when restricting to the real occurrences in the observed sample. In doing so, we avoid that the mentioned consecutive rank of an hapax is outside the set $\mathbf{Rank}$. Such a restriction will be removed in the next section, where we will deal with future occurrences and construct the data generating process.

Let us consider a discrete-time stochastic process $X=(X(t):t \in \mathbb{N})$ taking values in a set called $\mathbf{Rank}$ and $P$ the related probability law. $X$ is a Markov chain of order one if we have
\begin{equation}
\label{ord1}
P(X(t+1)=i_{t+1}|X(t)=i_t)=P(X(t+1)=i_{t+1}|X(t)=i_t, \dots, X(1)=i_1, X(0)=i_0),
\end{equation}
for each $t\in \mathbb{N}$ and $i_0, i_1, \dots, i_t, i_{t+1} \in \mathbf{Rank}$.

In the context of empirical studies, the transition probabilities can be estimated based on the observed transition frequencies. However, the empirical estimation of Eq. (\ref{ord1})'s second term offers a high level of computational complexity -- as also acknowledged by \cite{Renner, Friedrich}. Therefore -- and in line with the quoted papers -- we can write a less heavy relationship and say that $X$ is of order one when the following simplification of Eq. (\ref{ord1}) is true:
\begin{equation}
\label{ord2}
P(X(t+1)=i_{t+1}|X(t)=i_t)=P(X(t+1)=i_{t+1}|X(t)=i_t, X(t-1)=i_{t-1}),
\end{equation}
for each $t\in \mathbb{N}$ and $i_{t-1}, i_t, i_{t+1} \in \mathbf{Rank}$.

So, the transition probabilities of orders one and two appearing in Eq. (\ref{ord2}) have been empirically estimated by looking at the consecutive empirical observations.

Two steps are now carried out. In the \emph{First step}, we check the validity of Eq. (\ref{ord2}); in doing so, we test that the empirical transition probabilities estimated by the original sample are associated with a Markov chain of order one. In the \emph{Second step}, we show that the obtained Markov chain of order one is a statistically sounding representation of the original sample, hence leading to the final statement of the Markovianity of the data.

It is important to notice that the \emph{First step} could be enough to state the Markovianity of $X$. Indeed, by quoting \cite{Friedrich}, condition (\ref{ord2}) \emph{is a strong indication that the data set possesses Markovian properties}. However, we also present the additional analysis performed in the \emph{Second step} for providing more robust support to the Markovian probabilistic structure of the historical evolution of the ranks, having our empirical sample as an outcome.
\begin{itemize}
\item \emph{First step.}
\end{itemize}
We have run 1000 simulations for both order one and two transition matrices. This step originates 2000 samples from Markov chains. The lengths of the series calculated from the first- and second-order transition matrices are different. The series simulated with the former matrix have 509138 elements -- which is precisely the empirical sample's length, so we can also use such simulations for the \emph{Second step} (see below) -- while the ones simulated with the latter matrix have 100000 values.

\begin{figure}
\includegraphics[width=406.0pt]{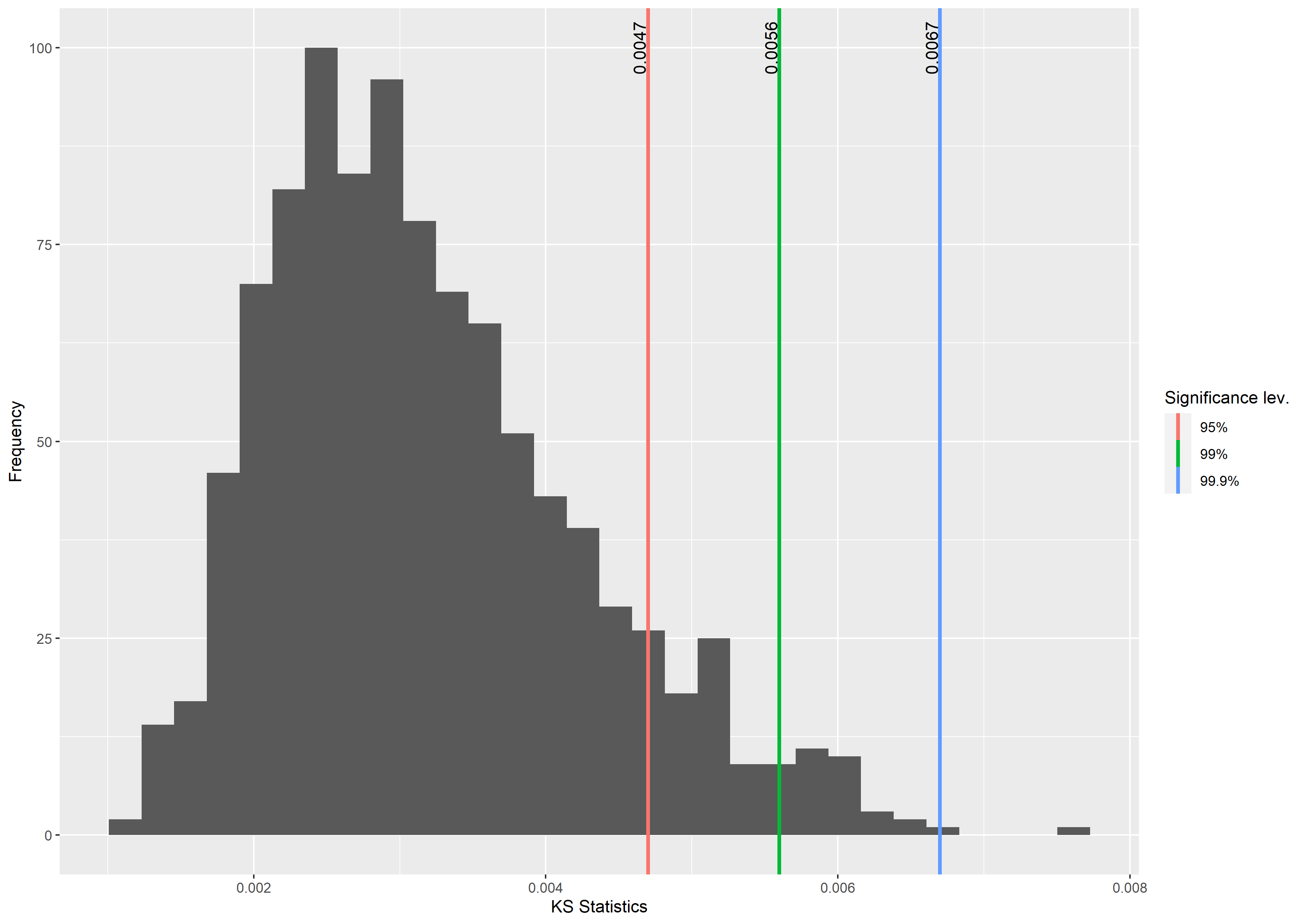}
\caption{The graph contains the Kolmogorov-Smirnov tests statistics generated by comparing the simulated series obtained using the empirical first-order and second-order transition probabilities. The coloured lines represent the thresholds calculated at various significance levels ($95\%,99\%,99.9\%$).}\label{KSfirstsecod}
\end{figure}

Then, the first- and second-order simulated series have been pairwise compared via the Kolmogorov-Smirnov (KS) test \citep[see, e.g.,][where the test has been employed to test if power laws fit binned empirical data]{Yogesh2014}. Namely, the empirical distributions of each simulated series from the first-order transition matrix have been compared with those of the simulated series coming from the second-order transition matrix. The KS statistics are reported in Figure \ref{KSfirstsecod} where we also report the thresholds of the statistics at different significance levels to make the reading more accessible. Almost all the series can be considered as coming from the same distribution; therefore, there is no statistical difference between the first-order transition probabilities and the second-order ones. This outcome supports the validity of Eq. (\ref{ord2}) so that the Markov chain obtained by using the empirical transition probabilities estimated from the original sample is of order one.

\begin{figure}
\includegraphics[width=406.0pt]{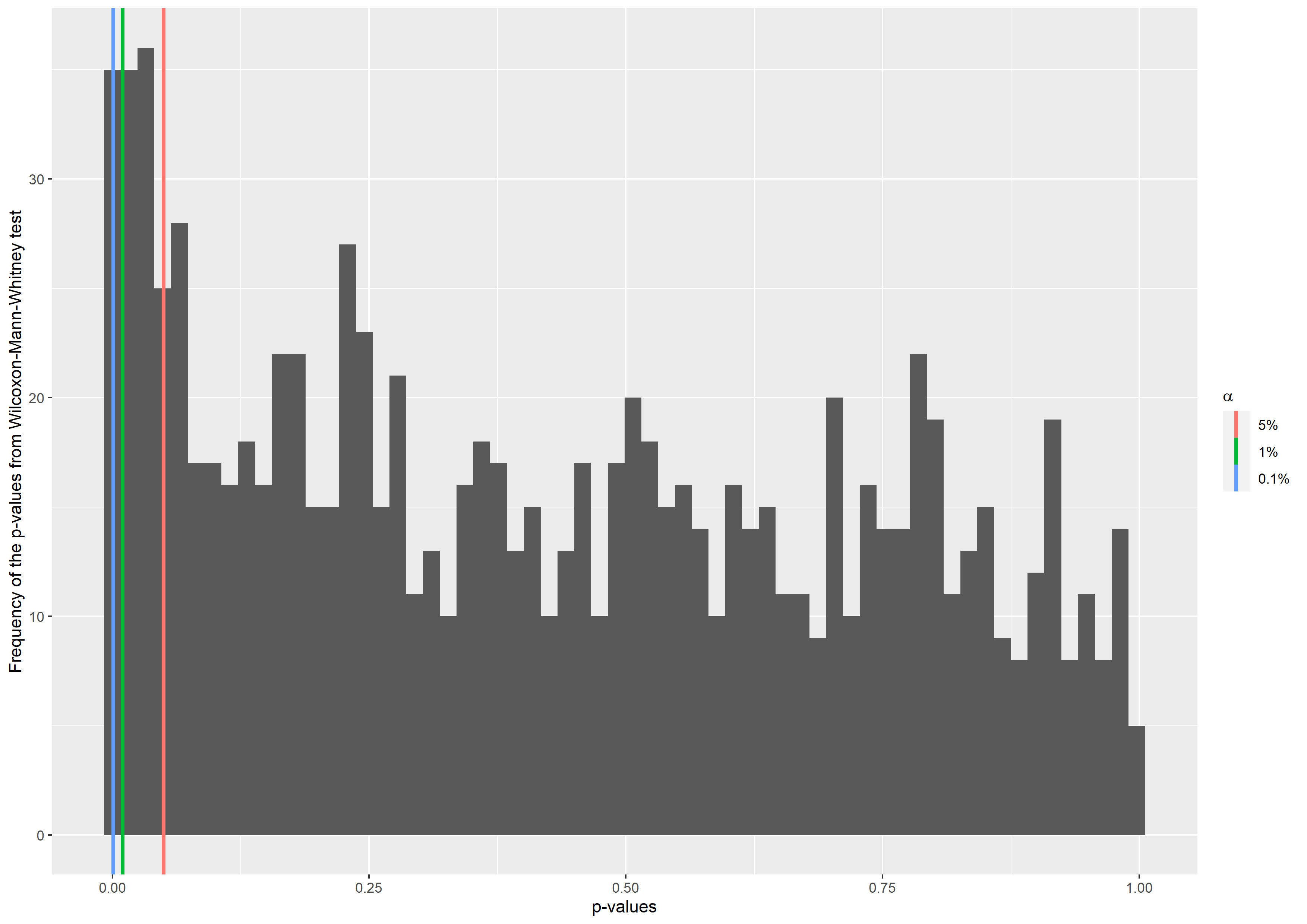}
\caption{The histogram contains the frequencies of Wilcoxon-Mann-Whitney tests' p-values generated by comparing the series simulated with the first and second-order transition matrix calibrated on the observed data. The coloured lines represent the thresholds  ($5\%,1\%,0.01\%$).}\label{WMWtest}
\end{figure}

To additionally verify the validity of Eq. (\ref{ord2}), we have compared the simulated series -- with the first- and second-order transition matrices -- with each other employing the Wilcoxon-Mann-Whitney test \citep[see][where the test has been applied in different fields but for similar purposes]{Zipunnikov2014,barry2008statistical}. Such a test is explicitly identified by \cite{Friedrich} as one of the methods to be used for checking Eq. (\ref{ord2}). The resulting p-values reported in Figure \ref{WMWtest} confirm the KS tests' results.
Indeed, by visually inspecting Figure \ref{WMWtest}, one can notice that most of the p-values -- about 90\% of them -- are larger than the $5\%$. Therefore, we have no compelling evidence to conclude that the series data differ and, rather than this, we can conclude that the series are obtained by the same data generating process. This finding strongly supports that the observed data are realisations of a first-order Markov chain.

\begin{itemize}
\item \emph{Second step.}
\end{itemize}
We now check that the Markov chain of order one with states space $\mathbf{Rank}$ and whose transition probabilities are estimated from the available observations can be seen as the data generating process of the original sample. In so doing, we provide additional evidence that the evolutive process related to the formation of the target distribution for our MCMC procedure has a Markovian structure.

We consider the 1000 sampling from the first-order Markov chain already created in the \emph{First step}. We recall that the length of each simulated series is the one of the observed empirical sample, i.e. it is 509138.

We have compared the simulated series to the original one from several perspectives.

\begin{figure}
\includegraphics[width=406.0pt]{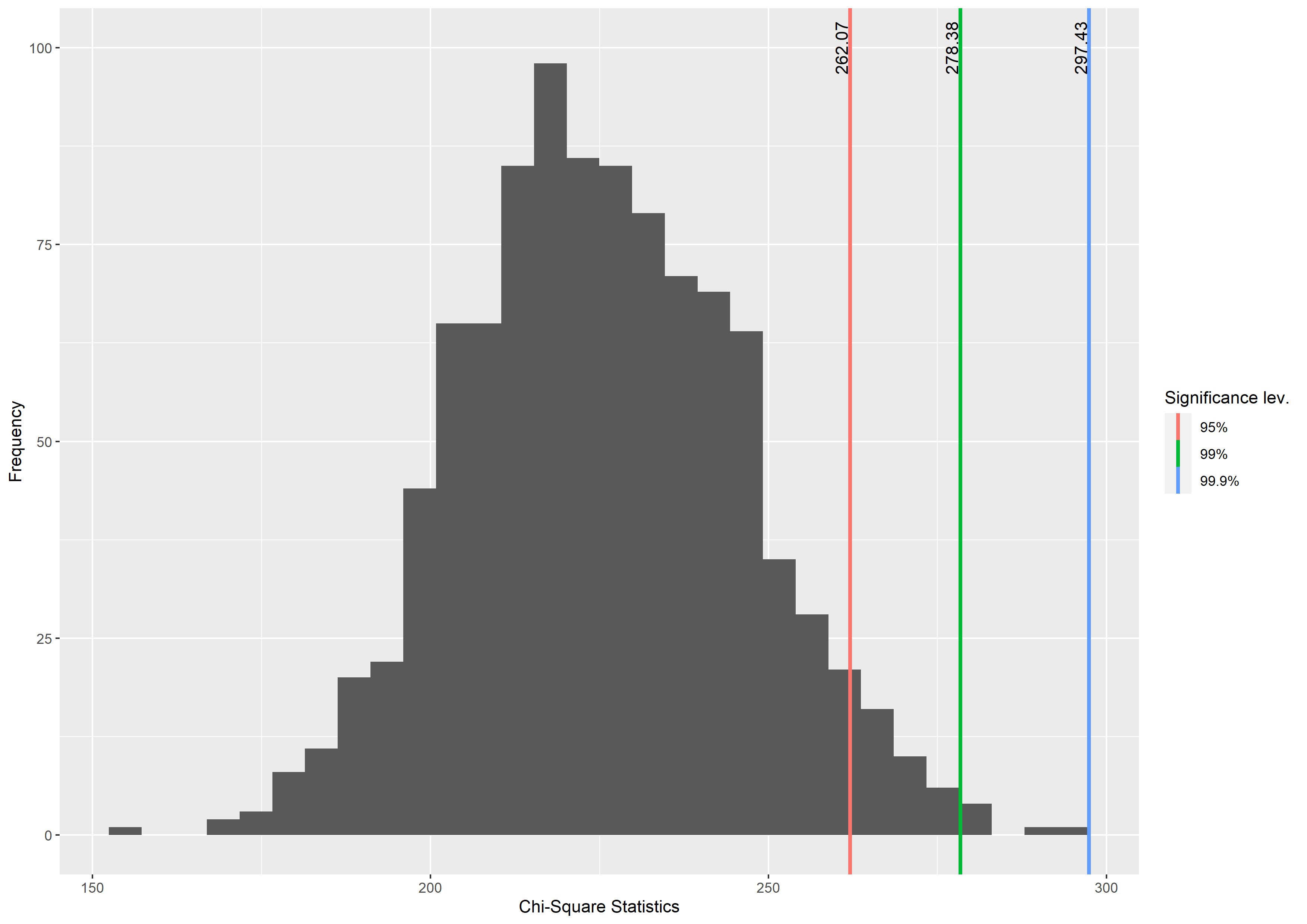}
\caption{The graph contains the Chi-square tests statistics generated by comparing the empirical distribution of the ranks from the observed hapax series and the empirical distributions coming from the simulated series. The degrees of freedom are 226, which is the cardinality of $\textbf{Rank}$ minus one. The coloured lines represent the thresholds calculated at various significance levels ($95\%,99\%,99.9\%$).}
\label{CS}
\end{figure}

First, the distribution of the values of each simulated series has been compared to the one obtained from the empirical ranked hapaxes through the Chi-square tests. For the majority of the cases, we accepted the null hypothesis at the $95\%$ significance level (see Figure \ref{CS}).

\begin{figure}
\includegraphics[width=406.0pt]{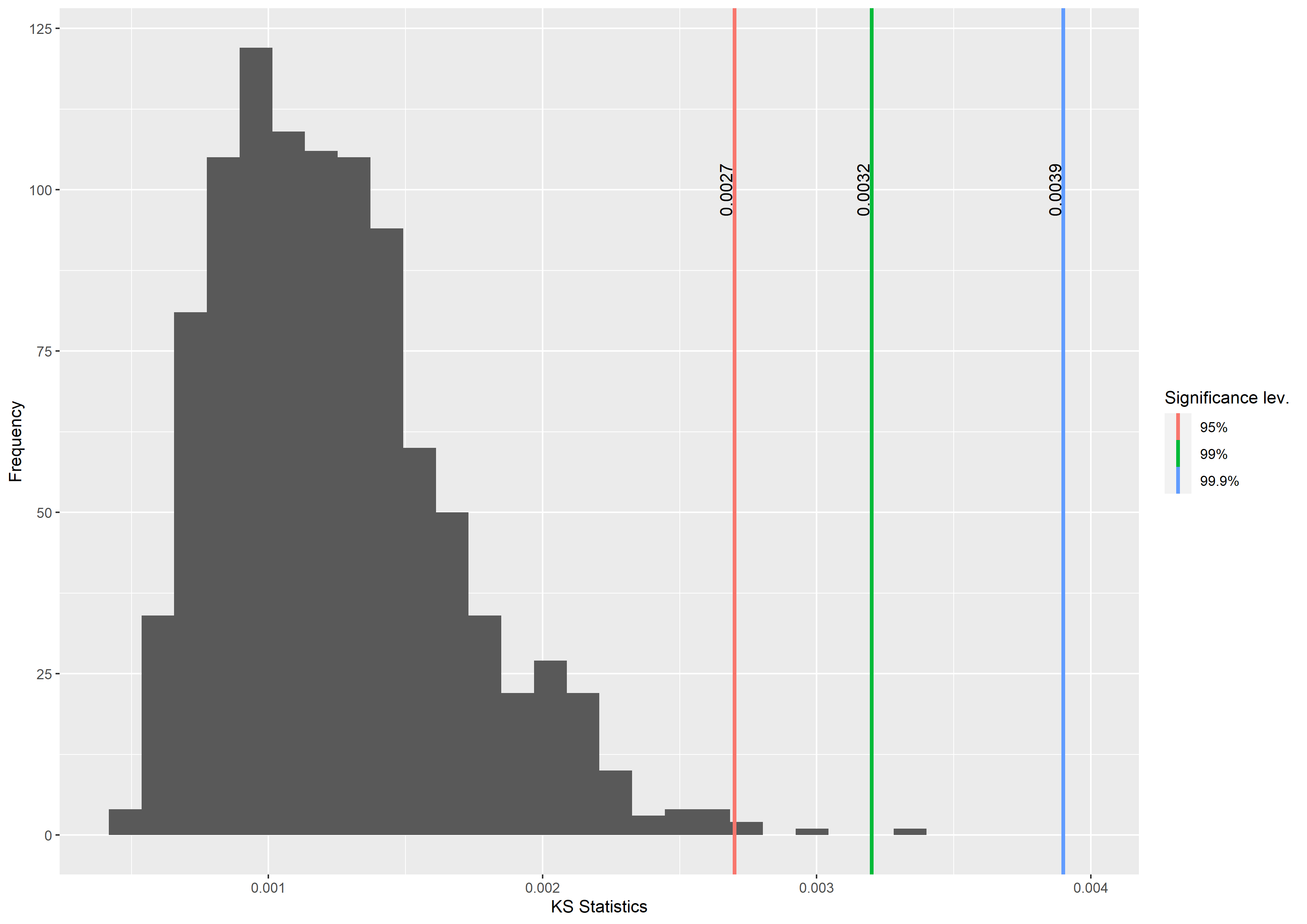}
\caption{The graph contains the Kolmogorov-Smirnov tests statistics generated by comparing the empirical distribution of the ranks from the observed hapax series and the empirical distributions coming from the simulated series. The coloured lines represent the thresholds calculated at various significance levels ($95\%,99\%,99.9\%$).}\label{KS}
\end{figure}

As a second device for comparing the distributions of the simulated series to the original one, we have implemented a Kolmogorov-Smirnov test. The obtained results are in line with the ones obtained for the Chi-square test. Furthermore, they seem to provide more convincing evidence of the validity of the tested condition (see Figure \ref{KS}).

\begin{figure}
\includegraphics[width=365.4pt]{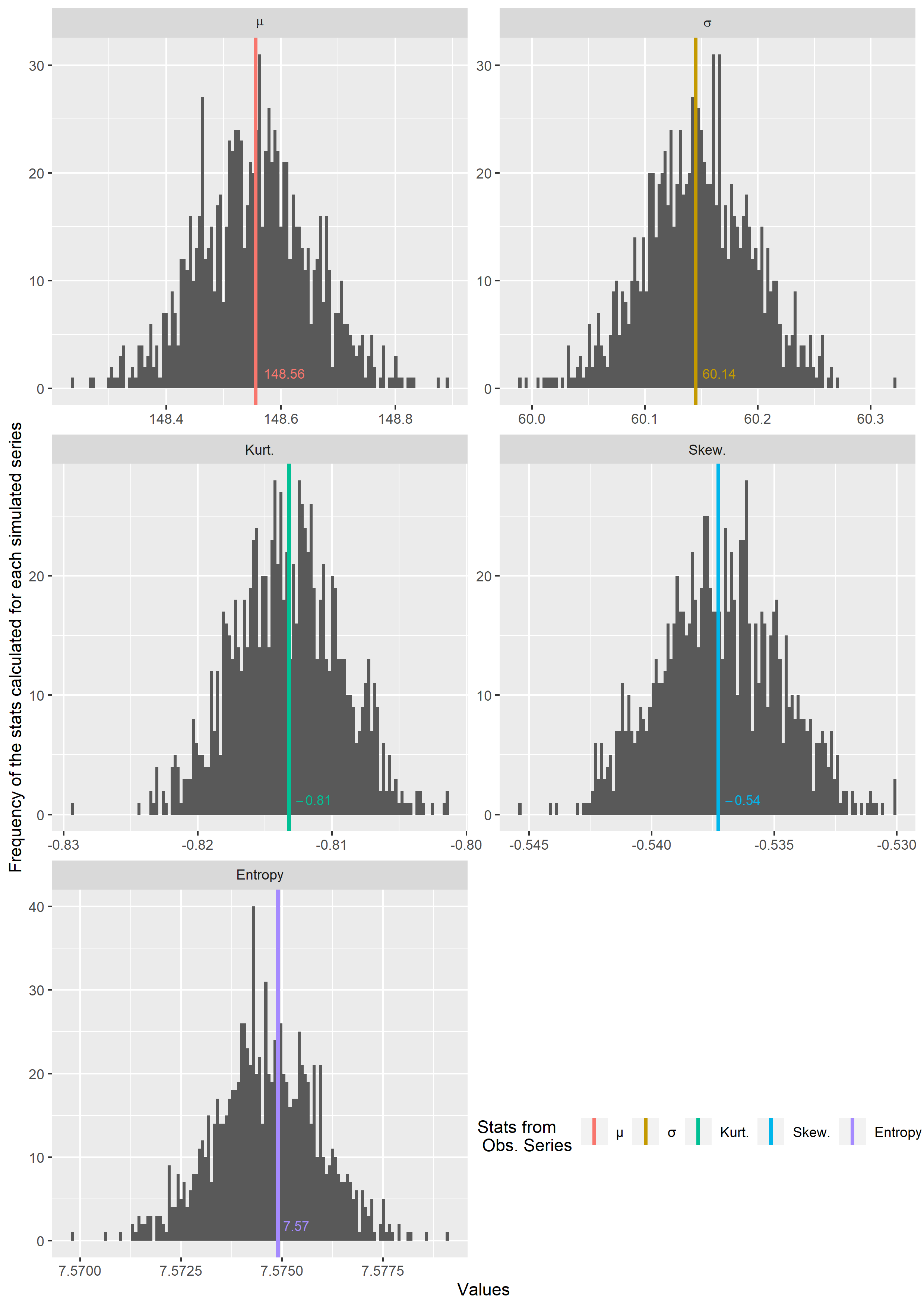}
\caption{Each sub-graph contains the distribution of the respective statistical indicator calculated for each simulated time series. The coloured lines represent the values of the same indicators coming from the observed time series of hapaxes' ranks.}\label{stats}
\end{figure}

To additionally verify that the hapaxes series is a realization of the considered first-order Markov chains, we have calculated some relevant statistical indicators for the 1000 simulated series. We have compared them to those coming from the observed hapax series. We considered Mean ($\mu$), Standard Deviation ($\sigma$), Kurtosis, Skewness, and Shannon entropy. Figure \ref{stats} contains the results of such an analysis. By visually inspecting the outcomes, one can notice that the indicators' distributions from the simulated series are centred on the associated statistical indicators computed for the empirical sample.

To conclude, we have found broad statistical evidence that the considered sample of the ranks of the observed hapaxes can be viewed as a realization of a Markov chain of order one.

\section{Construction of the Markov chain generating the ranked data}
\label{stoch}

This section provides an algorithmic construction of a Markov chain with the observed ranked data as stationary distribution. As we will see, we follow a Markov Chain Monte Carlo (MCMC) approach.

We employ and adapt the same notation used in the previous sections with a reasonable abuse of notation.

The rank-size analysis performed in \cite{JOI} and recalled in Section \ref{JOI} assumes that the size of an hapax is given by its absolute frequency, namely by the number of times in which such a word has been pronounced once in a US President speech. Differently, we consider in this section the relative frequencies of the hapaxes. In particular, we construct the random variable $\mathcal{R}$ whose realisations are the ranks and accordingly to the Zipf-Mandelbrot law in Eq. (\ref{zipf}) with calibrated parameters as in Table \ref{Table1}. To this aim, we first set

\begin{equation}
\label{rank} \mathbb{P}(\mathcal{R}=r)=F_r, \qquad \forall r=1,
\dots, \bar{r},
\end{equation}
where $\bar{r}$ is a prefixed threshold \textit{larger enough} than
the empirically observed highest value for the rank and $F_r $ is
given by
\begin{equation}
\label{fr} F_r = \frac{f(r)}{\sum_{h=1}^{\bar{r}}f(h)}, \qquad
r=1,\dots, \bar{r}
\end{equation}
being $f(r)$ defined as in Eq. (\ref{zipf}) with calibrated parameters
as in Table \ref{Table1}.

Equations (\ref{zipf}) and (\ref{fr}) assure that $F_r$ decreases
with respect to $r$. The criteria to select the value of $\bar{r}$
-- and the meaning of the term \textit{larger enough} mentioned
above -- will be discussed later.

Evidently, the vector $F=(F_1, \dots, F_{\bar{r}})$ represents a discrete probability distribution over the set $\mathbf{Rank}=\{1,\dots, \bar{r}\}$ collecting the ranks. {As we will see in detail below, $F$ is the stationary (target) distribution of the MCMC procedure.}

We here construct a Markov chain in the light of the \textbf{Problem P} introduced in Section \ref{IIInew}.

The formulation of \textbf{Problem P} gives the intuition of the meaning of \textit{large enough} associated to $\bar{r}$. Indeed, we here project the analysis in the future and do not restrict here only to the set of observations. Thus, the ranks of future hapaxes should also include the possibility of going beyond the maximum rank of the original sample, and a large value of $\bar{r}$ guarantees a consistent procedure in this sense.

The \textit{temporal evolution} of the hapaxes has to be intended in the sense of the proposed overall analysis of the speeches, which
means that two hapaxes are consecutive when they are sequentially pronounced in the same speech, or one is the last of
a speech and its consecutive is the first of the following speech with at least one hapax.

We introduce a homogeneous Markov chain $X=(X(t):t \in \mathbb{N})$
with finite states space $\mathbf{Rank}$. We assume that hapaxes
are not evaluated in the light of their semantic values but through
their ranks, and they have been pronounced accordingly to the Markov chain $X$. In so doing, we propose a response to \textbf{Problem P} by
dealing with the conditioned probability
\begin{equation}
\label{P} \mathbb{P}(X(t+1)=j | X(t)=i) \qquad \forall\,i,j \in
\mathbf{Rank}.
\end{equation}

Since the Markov chain $X$ is taken homogeneous, then
\textbf{Problem P} reduces to the identification of the $\bar{r}
\times \bar{r}$ transition probability matrix $\Pi=(\pi_{ij}:i,j \in
\mathbf{Rank})$, where $\pi_{ij}=\mathbb{P}(X(1)=j | X(0)=i)$ is the
element at row $i$ and column $j$ (again, where $i$ and $j$ also represent ranks). Namely, we are considering the probability of getting an hapax of rank $j$ after that a hapax of rank $i$ has been employed by the speaker.

To identify $\Pi$, we assume that the probability distribution in $F$ represents the stationary distribution of the Markov chain. Such an assumption lies in the evidence that the original sample of the entire set of the Presidents' speeches is so large that $F$ may be effectively viewed as the limiting distribution of the Markov chain when time goes to infinity.

This said the \textbf{Problem P} can be mathematically translated to
the following stochastic calculus problem.

\begin{itemize}
\item[\textbf{Problem P'}] \textit{Identify
a squared stochastic probability matrix
$\Pi$ of order $\bar{r}$ such that the homogeneous Markov chain $X
=( X(t) : t \in \mathbb{N})$ with $\Pi$ as transition matrix has
stationary distribution $F$.}%
\end{itemize}

Standard Markov chain theory gives conditions for the existence and
the uniqueness of the stationary distribution of a given Markov chain.
We are dealing here with the inverse problem of assessing the transition probabilities from the knowledge of the limiting distribution. We cannot expect the existence of a unique Markov chain with state-space $\mathbf{Rank}$ and stationary distribution $F$.

The relationship between the stationary distribution $F$ and the
transition probability matrix $\Pi$, along with the explicit reference to the stochasticity condition of $\Pi$, can be written in matrix form as follows:

\begin{equation}
\label{solP} \left\{
  \begin{array}{l}
    F \cdot \Pi = F \\
    \Pi \cdot \mathbf{1}(\bar{r},1)=\mathbf{1}(\bar{r},1)  \end{array}
\right.
\end{equation}
where $\mathbf{1}(\bar{r},1)$ is the $\bar{r}$-dimensional column
vector filled by ones.

Eq. (\ref{solP}) corresponds to a linear system with $2\bar{r}$
equations in $\bar{r} \times \bar{r}$ variables, so that it cannot have a unique solution in the general case of $\bar{r}\geq 3$ -- which is exactly the cases we are dealing with.

More specifically, the classical Rouch\'{e}-Capelli Theorem states that one of the following alternatives is true: the system does not admit solutions, or it can have an infinite number of solutions dependently at least on $\bar{r}^2-2\bar{r}$ parameters. In both cases, the assessment of the transition probability matrix through the closed-form expression in Eq. (\ref{solP}) does not represent an efficient way to solve \textbf{Problem P'}.

Thus, one can reasonably deal with a numerical rewriting of
\textbf{Problem P'}, whose algorithmic version can be rewritten as it
follows:
\begin{itemize}
\item[\textbf{Problem P''}] \textit{Identify a homogeneous Markov chain $X
=( X(t) : t \in \mathbb{N})$ with state space $\mathbf{Rank}$ and
whose empirical distribution converges to $F$ as $t$ goes to
infinity.}
\end{itemize}
In formula, \textbf{Problem P''} means that we are looking for a
Markov chain $X$ such that
 \begin{equation}\label{P''}
\mathbb{P}\left(\lim_{t \to \infty}  \frac{1}{t} \sum_{s=0}^{t-1}
\mathbf{1}_{\{ X(s)=r \}} = F_r\right)=1  , \qquad \forall\, r \in
\mathbf{Rank},
\end{equation}
where $\mathbf{1}_{\bullet}$ is the indicator function of set
$\bullet$.

The assessment of the Markov chain $X$ which satisfies the condition (\ref{P''}) is pursued by adopting an MCMC procedure, with a specific focus on the Metropolis-Hastings algorithm (for the details of this numerical strategy, refer to the original contributions \citealt{metropolis, hastings}). The implementation details are reported in the next subsection.

Indeed, the scope of an MCMC procedure is to construct a reversible regular Markov chain with an \textit{a priori} given stationary distribution. Among the wide set of the proposed algorithms in the literature, Metropolis-Hastings is particularly appropriate. The only drawback of the Metropolis-Hastings algorithm appears when the state space of the Markov chain is huge. In our specific case, we do not have a critically large cardinality for the state space of the Markov chain, in that the most frequent hapax occurs 250 times \citep[and the maximum rank is 227 as also stated in][]{JOI}. Therefore, a reasonable value of $\bar{r}$ should be greater than 227 but not too far from such a value.

The Metropolis-Hastings procedure is based on an iterative generation of a sampled value $j$ at a time $t+1$ conditioned to the value $i$ at time $t$. The generation is based on a transition probability $q_{ij}$ which can be prefixed according to a specified conditional probability distribution over the set $\mathbf{Rank}$ -- specifically, a uniform distribution over the set $\mathbf{Rank}$ in our experiments.
Then, a function $a: \mathbf{Rank}^2 \to [0,1]$ is introduced, so
that $(i,j) \mapsto a(i,j)$, which is the probability of accepting
$j$ at time $t+1$ once $i$ has been observed at time $t$. Such a
probability does not depend on $t$, and it is given by
\begin{equation}
\label{aij}
a(i,j)=\min \left\{1,\frac{F_j q_{ji}}{F_i q_{ij}}\right\} \qquad
\forall\,i,j \in \mathbf{Rank}.
\end{equation}
If the value $j$ is rejected, then the process remains in $i$. In so doing, we construct iteratively the Markov chain $X$.

\subsection{Experimental test and convergence}
\label{subsec:convergence}
To test the validity of the stochastic process class proposed in the previous section, we implement the Metropolis-Hastings algorithm and here we describe the experimental setup employed.

In our simulation procedure, we assume uniform transition probabilities, so that $q_{ij}=1/\bar{r}$ for each $i,j \in \mathbf{Rank}$. In so doing, Eq. \eqref{aij} becomes:
\begin{equation}
\label{aj}
a(i,j)=\min \left\{1, \frac{F_j}{F_{i}} \right\} \qquad \forall\,i,j \in \mathbf{Rank}.
\end{equation}

We set $\bar{r} = 300$. Furthermore, Eq. \eqref{fr} with the parameter setting presented in Table \ref{Table1} is used to generate the probability of extracting a certain rank. Given this framework, we need to test the algorithm's convergence to the target distribution. Indeed, convergence is a relevant issue to be explored since it is not true that the MCMC algorithm creates a Markov chain that inevitably converges to the target asymptotic distribution. We briefly elaborate on this point.

As already said, some critical aspects related to convergence can be found in the existence of tiny transition probabilities -- a typical feature of state spaces with large cardinality -- or in identifying the state space structure in the multivariate setting. The former criticism appears when the Metropolis-Hastings algorithm is used -- which is exactly the adopted procedure; the latter one is associated with the case of the Gibbs sampler \citep[see, e.g. ][]{Geman}. Interestingly, Diaconis proposed his own reading of how the convergence of the MCMC algorithms is a central theme in applied probability. In \cite{Diaconis9, Diaconis13}, this outstanding scientist traces the route for future developments of MCMC in both theoretical and application advancements.
A relevant discussion on the convergence of the MCMC procedures can be found in \cite{Sinharay}. The author presents a survey on the matter and acknowledges that \emph{"A number of other diagnostics compare the sampled distributions obtained from the MCMC for different runs [...]; convergence is concluded when the difference between some aspects of the empirical distributions [...] over the different runs is negligible in some sense"}. The quoted paper states also that \emph{[...] "one should measure the distance between the sample distribution and the target posterior distribution" [...]}. In this paper, we take for us these pieces of advice by measuring the distance -- KS, as we will see below -- between the sampled distributions for different runs and the posterior target distribution.

At this aim, we simulate the generation of $N = 100000$ consecutive hapaxes' ranks, and the results are saved in a vector $X$. The algorithm is implemented in R; here is a summary of it:
\\
\begin{algorithm}[H]
\tcc{for cycle to generate the $N$ step of the MCMC}
\For{$t \leftarrow 1$ \KwTo
$N-1$}{
 j $\sim$ Uniform$(\mathbf{Rank})$ \tcc*[r]{extr. of a number from a distribution which is uniform in $\mathbf{Rank}$}
 a $\leftarrow \min \left\{1,\dfrac{Fr(j) }{Fr(x_t)}\right\}$\;
 u $\sim$ Uniform$(0,1)$ \tcc*[r]{extr. of a number from a uniform in $(0,1)$ distribution}
 \eIf(\tcc*[f]{if to decide which is the step $x_{t+1}$}){$u \leq  a$}{$x_{t+1}\leftarrow j$\;}{$x_{t+1}\leftarrow x_{t}$\;}
} \caption{MCMC algorithm to generate the chain of hapaxes
ranks}\label{MCMC }
\end{algorithm}

\begin{figure}
\includegraphics[width=406.0pt]{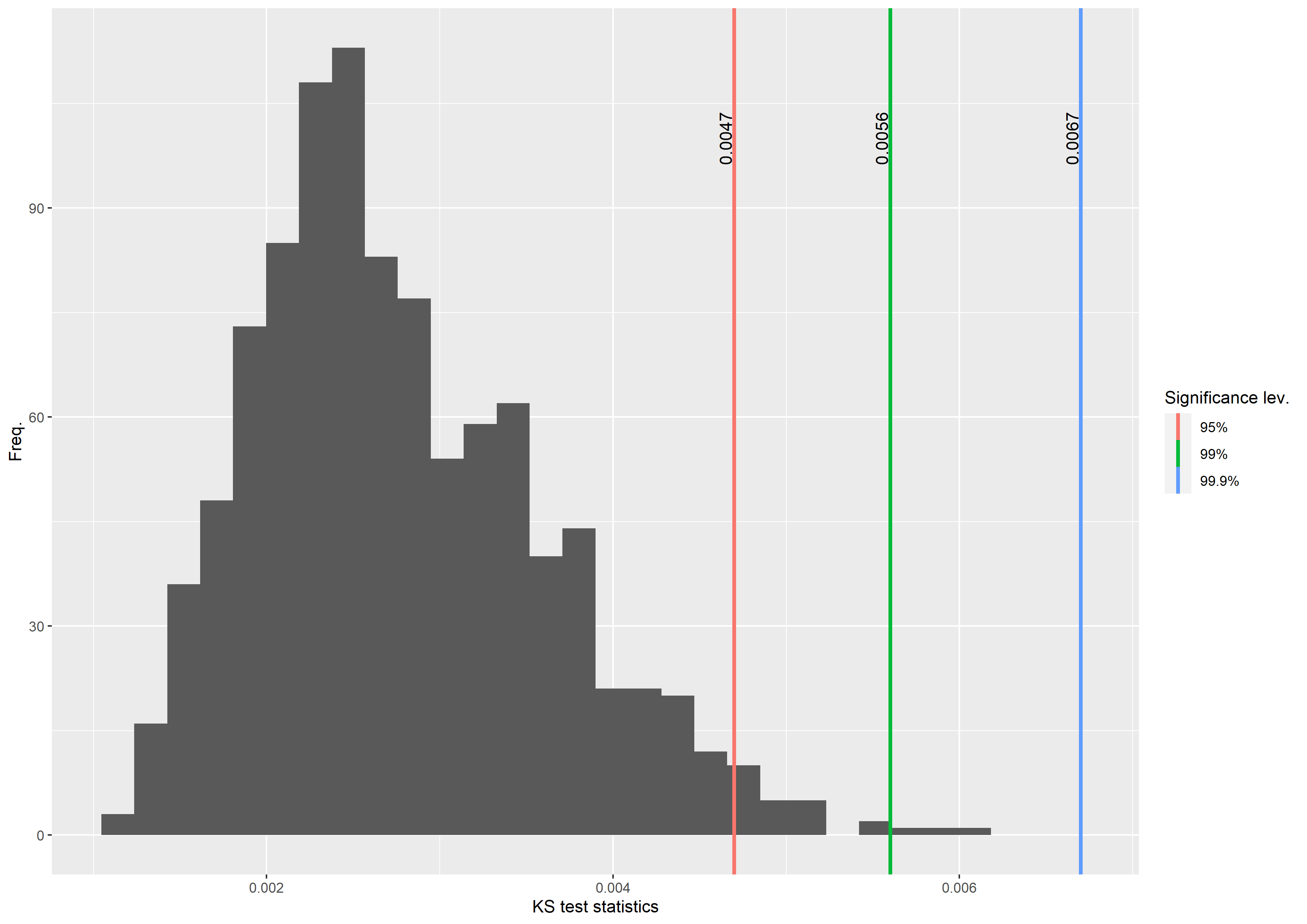}
\caption{Empirical distribution of the KS test statistics coming
form 1000 applications of the MCMC simulation procedure. The colored lines represent the thresholds calculated at various significance levels ($95\%,\,99\%,\,99.9\%$).}\label{dstathsit}
\end{figure}

The distribution of the $N$ simulated realisations of the Markov chain is then derived.
We then perform the KS test to obtain the statistical similarity between $F$ and the distribution obtained through the MCMC procedure. The KS test is implemented by computing the differences between the empirical
cumulative distribution and the simulated ones. The biggest absolute differences are taken as KS test statistics and compared to the threshold values calculated as in \cite{knuth1997} (Eq. (15) in Section 3.3.1), namely with the following formula:
\begin{equation}
\label{KStrs} T_{n,m} = \sqrt{-0.5\ln{(\alpha)}\dfrac{N+M}{NM}}
\qquad \alpha =0.05, 0.01, 0.001,
\end{equation}
where $N$ and $M$ are the sample sizes -- respectively 100000 and 31074 in our case. The resulting limit values are reported in Table \ref{SKtest}. The MCMC procedure and the KS test are implemented 1000 times. The KS test statistics distribution over the 1000 MCMC procedures is summarised in Figure \ref{dstathsit}. As it is possible to see from Figure \ref{dstathsit}, the null hypothesis can be accepted in almost all cases. Therefore, the empirical rank distributions generated from the MCMC proposed here are statistically similar to the original distribution of the hapaxes ranks. This represents scientific proof of the goodness of the theoretical model proposal.

\begin{table}[ht]
\centering
\begin{tabular}{r|rrr}
  \hline \hline
 C.I. & 95\% & 99\% & 99.9\% \\
  \hline
Thresholds & 0.004697& 0.005629& 0.006743\\
   \hline \hline
\end{tabular}
\caption{Above the thresholds here reported the null hypothesis of
the KS test has to be rejected. They are computed by using
Eq.\eqref{KStrs}.} \label{SKtest}
\end{table}

\section{Conclusions}\label{conclusioni}

This paper faces the challenging theme of identifying the stochastic structure for the text-based rank-size laws. We tailor the scientific proposal to a real-life rank-size context from the text analysis of the hapaxes in the US Presidents' speeches. In so doing, we are able to justify the selection of a stochastic process -- a Markov chain, in this context. In particular, we provide supporting arguments on the Markovian nature of the hapaxes generation. More than this, we give statistical evidence that the considered ranked data are generated from a first-order Markov chain.

The Markov chain generating the hapaxes is discussed from a purely theoretical perspective; it is explicitly constructed by applying an MCMC procedure based on the Metropolis-Hastings algorithm. Importantly, we show that the proposed algorithm converges, so the MCMC procedure leads to a Markov chain with the observed sample as stationary distribution.

The present study moves a step toward the study of the stochastic structure of the rank-size analysis coming from the linguistic field -- and of the rank-size analysis, in a more general context -- which is a relevant theme in information sciences and applied statistics literature. The specific context we deal with also contributes to the comprehension of the evolutive patterns of the messages delivered by the US Presidents.

The proposed methodological procedure is versatile at two different levels. On one side, one can apply the proposed MCMC algorithm to other rank-size contexts related to textual analysis, with particular attention to the hapaxes in carefully written texts collected in a corpus. On the other side, one can take a very different context and replicate the methodological flow in the considered case. First, one provides evidence of the type of stochastic process underlying the generation of the ranked data; second, one applies a calibration procedure for detecting the specific stochastic process by taking values in the observed ranks and having the rank-size distribution as long-run distribution.

\bibliographystyle{apalike} 
\bibliography{MCMC}       


\end{document}